\documentclass[12pt,preprint]{emulateapj}
\usepackage{natbib}
\usepackage{graphicx}
\usepackage{amsmath}
\usepackage{xcolor}
\usepackage{longtable}

\shorttitle{The Star-Planet Connection}
\shortauthors{Hinkel \& Unterborn}

\def\teff{{T$_{\text{eff}}$\,\,}}
\def\lg{{$\log$(g)\,}}

\def\gtaprx{ \mathrel{ \vcenter{
      \offinterlineskip \hbox{$>$}
      \kern 0.3ex \hbox{$\sim$}    } } }

\def\ltaprx{ \mathrel{ \vcenter{
      \offinterlineskip \hbox{$<$}
      \kern 0.3ex \hbox{$\sim$}    } } }
      
\def\aj{{AJ}}
\def\apj{{ApJ}}
\def\apjs{{ApJS}}
\def\apjl{{ApJL}}

\def\aap{{A\&A}}

\def\mnras{{MNRAS}}
\def\nat{{Nature}}

\def\icarus{{Icarus}}

\newcommand\blfootnote[1]{  \begingroup
  \renewcommand\thefootnote{}\footnote{#1}  \addtocounter{footnote}{-1}  \endgroup
}
   
\begin{document} 

\title{The Star-Planet Connection I: Using Stellar Composition to Observationally Constrain Planetary Mineralogy for the Ten Closest$^\dagger$ Stars}

\author{
  Natalie R. Hinkel\altaffilmark{1} and
  Cayman T. Unterborn\altaffilmark{2}
}
\email{natalie.hinkel@gmail.com}
\altaffiltext{1}{Department of Physics \& Astronomy, Vanderbilt University, Nashville, TN 37235, USA}
\altaffiltext{2}{School of Earth and Space Exploration, Arizona State University, Tempe, AZ, 85287, USA}

\begin{abstract}

The compositions of stars and planets are connected, yet, the definition of ``habitability" and the ``habitable zone" only take into account the physical relationship between the star and planet. Planets, however, are made truly habitable by both chemical and physical processes which regulate climatic and geochemical cycling between atmosphere, surface, and interior reservoirs. Despite this, ``Earth-like'' is often defined as a planet made of a mixture of rock and Fe that is roughly 1 Earth-density. To understand the interior of a terrestrial planet, the stellar abundances of planet-building elements (e.g. Mg, Si, and Fe) can be utilized as a proxy for the planet's composition.  We explore the planetary mineralogy and structure for fictive planets around the 10 closest stars to the Sun using stellar abundances from the Hypatia Catalog. Despite our sample containing stars both sub- and super-solar in their abundances, we find that the mineralogies are very similar for all 10 planets -- since the error or spread in the stellar abundances create significant degeneracy in the models. We show that abundance uncertainties need to be on the order of [Fe/H] $<$ 0.02 dex, [Si/H] $<$ 0.01 dex, [Al/H] $<$ 0.002 dex, while [Mg/H] and [Ca/H] $<$ 0.001 dex, in order to distinguish two unique planetary populations in our sample of 10 stars. While these precisions are high, we believe they are possible given certain abundance techniques, in addition to methodological transparency, recently demonstrated in the literature. However, without these precisions, the uncertainty in planetary structures will be so high that we will be unable to say confidently that a planet is like the Earth, or unlike anything we've ever seen.

\keywords{stars: abundances -- planetary systems -- planets and satellites: detection}

\end{abstract}

\section{Introduction}
Stars and planets are formed at roughly the same time and from the same original cloud of gas and dust. Whether the planets were formed due to core accretion \citep{Pollack:1996p7814} or gravitational instability \citep{Boss:1997p7829}, the composition of the star and planet are inextricably linked. Provided optimum geometry, it is possible to measure the composition of an exoplanet's atmosphere via the absorption of the host star's light. Unfortunately this technique is difficult given the atmospheric signal perturbation is very small, $\sim10^{-3}-10^{-5}$, with respect to the stellar spectra \citep{Demming17}. 
\blfootnote{\,\,\,$^\dagger$\,We made some cuts and ruled out a number of stars, but these 10 are still rather nearby.}

We are yet unable to directly observe the composition of a solid planetary regime, which is an important consideration when discussing ``habitable" and ``Earth-like'' planets \citep{Tasker17}. Instead, an observed planet's mass and radius are often used to constrain the solid, bulk composition of a planet: e.g. the relative size of its core and mantle \citep{Dorn15, Unterborn_2016}. However, the mass-radius relationship alone is limited in its utility for distinguishing fine detail for a terrestrial planet interior, due to the degeneracies inherent when bulk composition is left as a free parameter. As a result, we are only able to simply state that a planet is rocky or icy. Therefore, stellar elemental abundance ratios are needed to break the mass-radius degeneracy and adopt a more holistic mass-radius-composition understanding of terrestrial exoplanets. 

A planet's host star composition, however, represents a currently underutilized indirect measurement of terrestrial planet composition. \citet{Thiabaud15} explored the elemental relationship between a star and a planet, namely whether composition is determined via chemical equilibrium (minimizing Gibbs free energy, \citealt{Elser12}) or whether the Fe/Si and Mg/Si are the same as the host star's composition. They concentrated on Mg, Si, and Fe, which are prone to being condensed into solid, rocky material since their condensation temperatures (T$_C$) are high. Elements with T$_C \gtaprx$ 1200 K are considered refractory while those with 900 K $\ltaprx$ T$_C \ltaprx$ 1200 K are moderately volatile. Additionally, \citet{Thiabaud15} analyzed C and O, or elements that are typically found in gases with T$_C$ $<$ $\sim$900 K, or volatiles. By employing a planet formation and composition model, while cycling through a range of abundance ratios, they were able to determine that elements present within a planet were identical to those within a star. Other recent work shows the Solar composition is a good proxy for the Earth's  and potentially Venus' bulk composition using self-consistent mass-radius relations \citep{Unterborn_2016}. 

Terrestrial planets are built mostly via refractory elements, namely Mg, Si, Fe and O. 
In other words, 95\% of the structure within the bulk Earth (namely the silicate-shell containing the mantle and crust) can be created using a combination of only these four elements \citep{McD03}.

The relative fractions of refractory elements, namely Si/Mg, affect the relative proportions of olivine to pyroxene at pressures and temperatures indicative of the upper mantle, as well as  Mg-perovskite (bridgmanite is the phase bearing both Mg and Fe) and ferropericlase in the lower mantle. These mineral distributions, in turn, can affect melting relations, the viscous and elastic (or viscoelastic) properties of the planet, and potential for long-term sustentation of tectonic processes \citep[][ and references therein]{Foley16, Unte17b}. Additionally, Fe/Mg is shown to affect the relative size of a planet's core, which in turn may alter the heat flow from the core into the mantle and subsequently out of the mantle at the surface. The prevalence of moderate volatiles can influence exoplanet crustal composition and affect tectonic processes by preferentially fractionating and creating more andesitic conditions, or the interaction between converging plates which lead to continental crust formation. These conditions play a key role in geochemical cycling and buoyancy of subducting material, which governs the sideways and downward movement of tectonic plates \citep{Unte17b}. 

Volatiles also play an important role. For example, planets with elevated C abundances are likely to contain reduced diamond in their mantles, limiting convective forces due to the diamond's increased viscosity and thermal conductivity \citep{Unterborn14}. As such, planet composition can drastically affect the geodynamic and geochemical processes present on a terrestrial exoplanet. Therefore, stellar composition provides us with a key, yet underused, observable for understanding the geologic nature of an exoplanetary system, and must be considered along with mass-radius studies, when describing planets as ``Earth-like". 

Therefore, if we are to define a more holistic view of ``habitability" and ``Earth-like,'' all of the available physical and chemical data needs to be utilized in order to understand planets at a level meaningful in our search for other Earths and even life. This is especially important with the upcoming missions focusing on characterizing exoplanets, i.e. the Transiting Exoplanet Survey Satellite (TESS), the CHaracterizing ExOPlanets Satellite (CHEOPS), the PLAnetary Transits and Oscillations of stars (PLATO) mission, and the Wide-Field InfraRed Survey Telescope (WFIRST). In this paper, we look at the 10 closest$^{\dagger}$ stars to the Sun and analyze their stellar abundances per the Hypatia Catalog \citep{Hinkel14}, as outlined in Section \ref{sample}. In Section \ref{exoplex}, we briefly assess the mineralogies of fictive planets around these stars (explained in more detail in Paper II, \citealt{UnterbornHinkel}), which are similar due to the errors associated with the stellar abundances. In Section \ref{chem}, we present an algebraic walk-through to determine the abundance precision needed to distinguish two unique populations of planet mineralogies, both in our sample of 10 and within the Hypatia Catalog as a whole. In Section \ref{physical}, we discuss what is physically known about the sample of 10 stars, such that any detected terrestrial planets can be quickly and easily characterized. Finally, we summarize our results in Section \ref{conc}.

\section{Sample Selection}\label{sample}

For our sample of targets, we turn to the place that is most often observed by astronomers and most interesting for habitability: the stars closest to the Sun. From a technological standpoint, nearby stars will have higher measurement precision for the stellar properties and therefore will be more closely monitored for orbiting planets. Because of their proximity, they will also have more high resolution stellar abundance measurements and for a wider variety of elements. As a result, the properties of discovered planets will be very well known, as can be seen for Proxima Centauri b \citep{Anglada16}.

To begin, we adopt to the Hypatia Catalog \citep{Hinkel14,Hinkel16,Hinkel17} which contains stellar abundances for FGK-type stars within 150 pc of the Sun\footnote{www.hypatiacatalog.com}. Stellar abundances are defined such that an element ratio in a star (*) is compared to that same element ratio in the Sun ($\odot$):
\begin{equation}
[X/H] = \log \,(
N_{X} / N_Y)_{*} - \, \log \,( N_{X} / N_Y )_{\odot}\,\,,
\end{equation}\label{dex}
where the square brackets indicate solar normalization and $N_{X}$ and $N_H$ are the number of X and Y atoms (in mol) per unit volume, respectively. The associated unit is a logarithmic unit: dex. Hypatia is an amalgamate dataset compiled of abundance measurements from $>$200 literature sources. It is the largest dataset of main-sequence stellar abundances for stars near to the Sun and currently boasts $\sim$28,000 abundance measurements for $\sim$6000 stars. Additionally, it is unbiased in it's inclusion. Namely, if a literature data set contains the abundance determination for Fe and one other element within main-sequence stars in the solar neighborhood (150 pc), then it is added into Hypatia. All datasets have their intrinsic solar normalization scale removed and replaced with \citet{Lodders:2009p3091} so that all abundance data will be on the same baseline. Different solar normalizations can cause a shift of 0.06 dex, on average, for all of [X/H] abundances \citep{Hinkel14}. We use Hypatia not only because of its breadth, but also because it allows a true understanding of how different groups using a variety techniques measure elemental abundances within a star, or the $spread$.

We searched the Hypatia Catalog for stars that are nearest to the Sun that also have abundance measurements for all five elements: Mg, Al, Si, Ca, and Fe. However, we did not wish to include gravitational or spectroscopic binaries in order to minimize overlap of the individual star's spectra. We have listed those binary stars not included in our analysis in the Appendix. The one exception that we made was with respect to HIP 108870 ($\epsilon$ Ind) that has a wide orbital brown dwarf at a distance of $>$1400 AU \citep{Scholz03}; note Pluto is at a distance of $\sim$40 AU from the Sun (see Section \ref{physical}). Additionally, we opted to exclude any nearby stellar systems that already had detected planets\footnote{exoplanetarchive.ipac.caltech.edu} -- also listed in the Appendix. In this way, we are able analyze systems with simple, theoretical planetary formation mechanisms, that don't involve planetary migration or fractionation of the stellar abundances with respect to composition. 

One of the most powerful tools within the Hypatia Catalog is the \textit{spread} in the abundance data. When multiple literature sources measure the same element within the same star, the $spread$ is defined as the range in those abundance determinations \citep[after they are renormalized to][]{Lodders:2009p3091}. In many instances, the $spread$ is larger than the quoted error from a given literature source, which reveals how truly ``well understood" abundance measurements can be within a star \citep[see][particularly the top of Figure 3, for a more thorough discussion]{Hinkel14}. In this vein, we found that a number of stars had abundance $spreads$ that were $>$ 0.70 dex for one or many of the element abundances. We found that the large spread, even when using the median values, did not give us reliable abundance values. Therefore, we retained stars that had only had abundance $spreads$ $<$ 0.50 dex -- those that were removed are listed in the Appendix.

\begin{figure}
\begin{center}
 \centerline{\includegraphics[width=10cm]{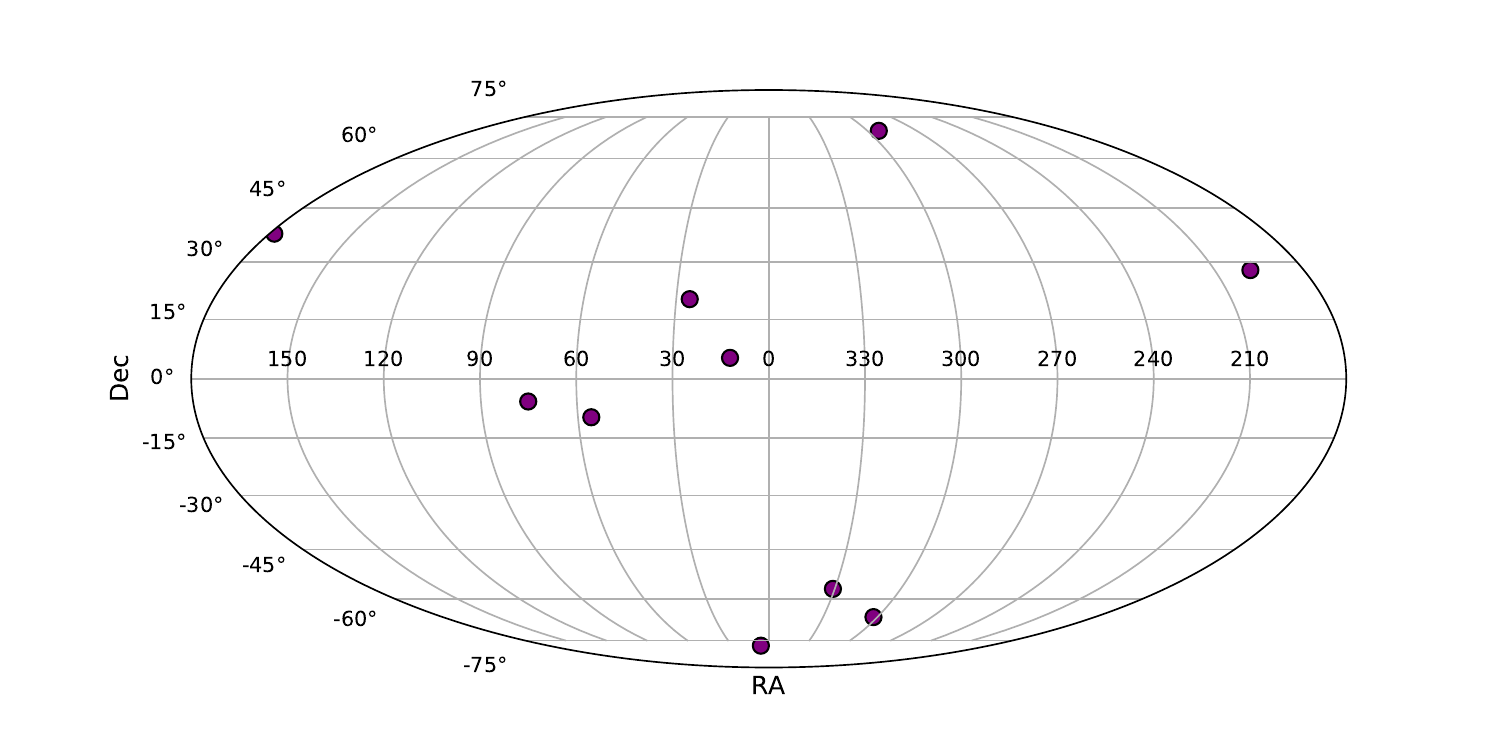}}
\end{center}
  \caption{Mollweide projection showing the sample of targets (see Table \ref{phys}).}
  \label{moll}
\end{figure}

We therefore define a sample of the 10 ``closest'' stars for the purposes of our study, which are listed in Table \ref{phys} with physical properties included from the Hypatia Catalog \citep{Hinkel17}. Of note, stellar distances vary from 3.62--7.53pc. The optimistic (Opt) and conservative (Cons) habitable zone (HZ) distances were determined using the calculator offered by the Virtual Planetary Laboratory\footnote{http://depts.washington.edu/naivpl/content/hz-calculator} per the equations in \citet{Kopp13,Kopp14}. Additionally, the location on the sky for the 10 stars is shown on a Mollweide projection in Figure \ref{moll}, where we see the large RA and Dec space covered by our selected sample.

\subsection{Stellar Abundances for the Sample}\label{sample-abunds}

The adopted stellar abundances from Hypatia for the sample of targets are given in Tables \ref{abunds1} and \ref{abunds2}. Per the Hypatia methodology \citep{Hinkel14}, the median value of measurements (as determined by multiple groups) are listed as [X/H] while the $spread$ or range of those values are given as sp[X/H]. The only exceptions are for HIP 73148 and HIP 99461 where the [Mg/H] abundance was measured by one group \citep[][respectively]{Luck:2005p1439,Maldonado15} and therefore had a spread = 0.0. For these two cases, we used the individual error for [Mg/H] within each star as reported directly by the literature source. Additionally, within Tables \ref{abunds1} and \ref{abunds2}, we have listed the [Mg/H] abundances associated with each element. The ExoPlex code (see Section \ref{exoplex}) requires that the molar ratios of the four elements (Ca, Al, Si, Fe) are with respect to Mg (see Section \ref{exoplex} and Table \ref{ratios}). The abundance and spread for [Mg/H]$_\text{Y}$ means that only those data sets that measured $both$ [Y/H] and [Mg/H] were used to calculate the final [Mg/H]$_\text{Y}$ value. In this way, we are able to ensure that we were comparing similar quantities from like-sources, as opposed to taking the median value of, for example, 7 datasets that measured [Fe/H] while only 2 who also measured [Mg/H]. Therefore, a unique [Mg/H] calculation was needed for each of the four elements. 

\begin{figure*}
\centering 
\includegraphics[height=5cm]{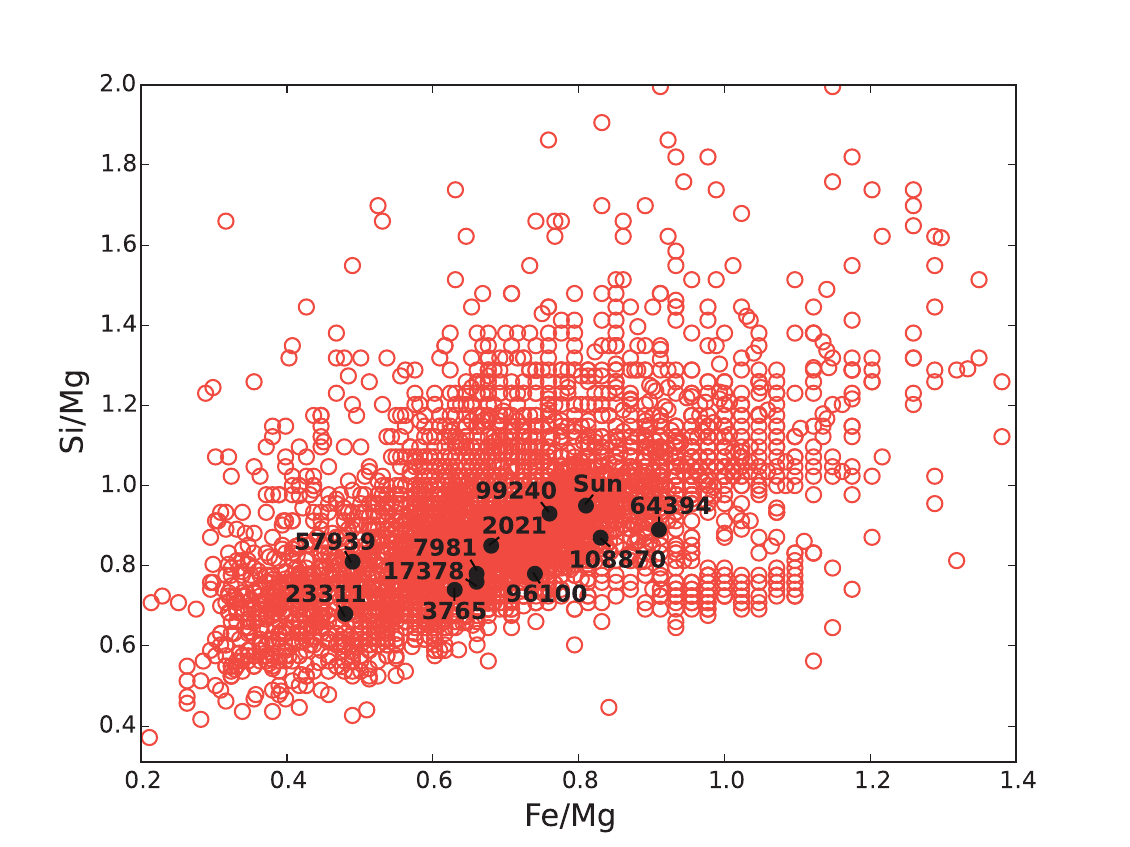}\quad
\includegraphics[height=5cm]{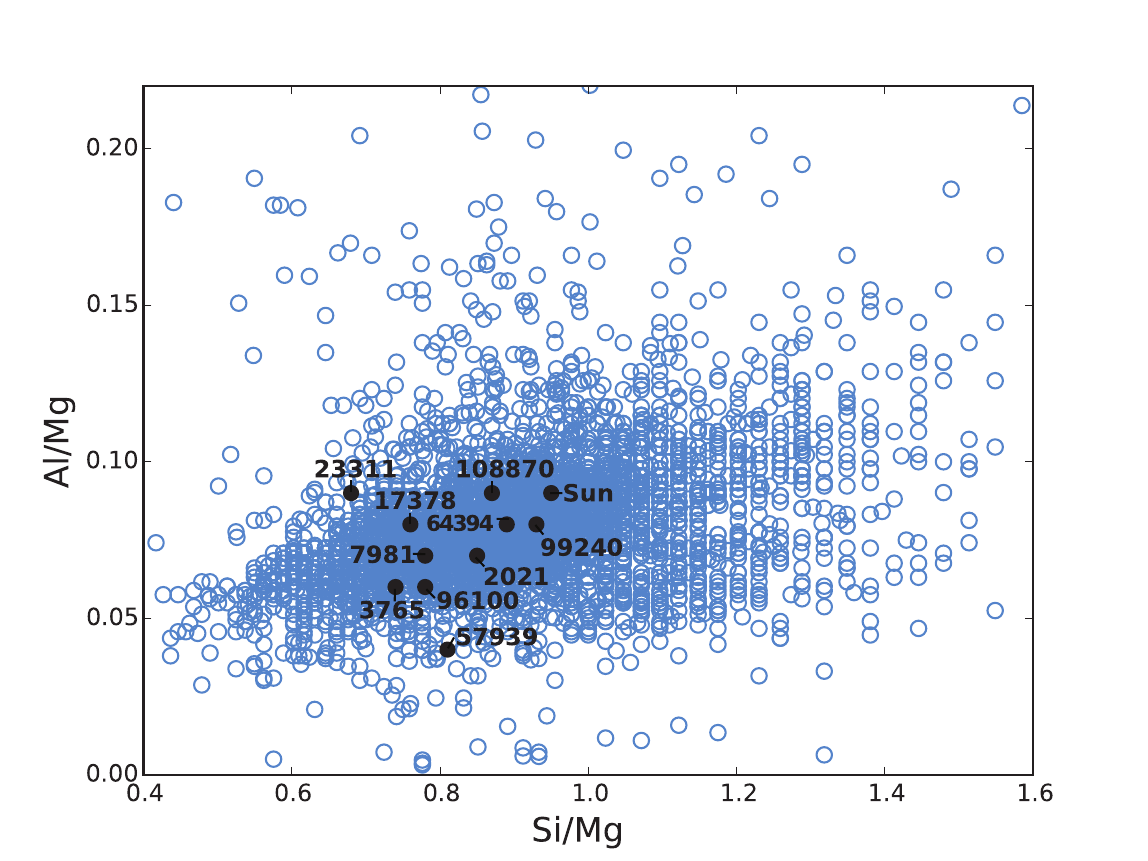}\par\medskip
\includegraphics[height=5.5cm]{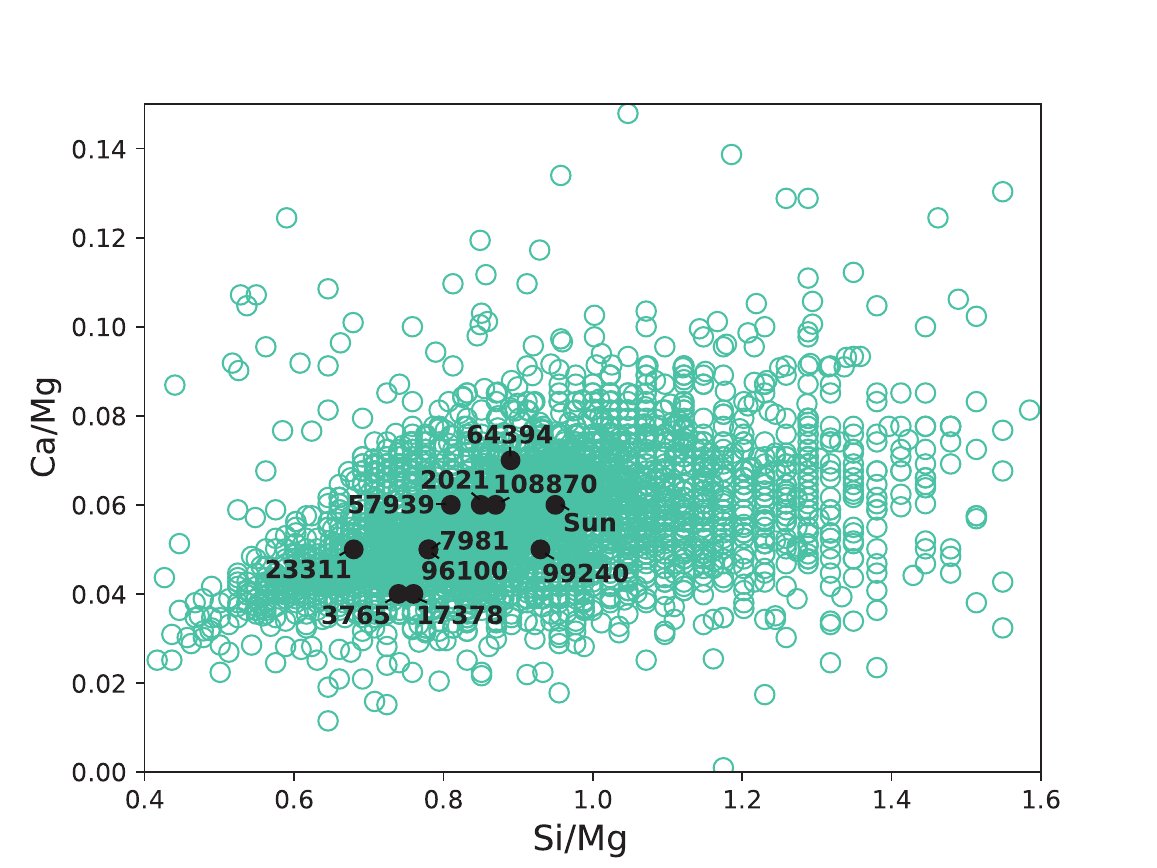}
\caption{Molar ratios for the sample of 10 stars, including the Sun, labelled. Left: Fe/Mg with respect to Si/Mg, where the Hypatia Catalog stars are in orange. Right: Si/Mg versus Al/Mg, with the Hypatia stars in blue. Bottom: Si/Mg with respect to Ca/Mg, where Hypatia stars are in light green.}
\label{SiMg}
\end{figure*}

In Figure \ref{SiMg} (left), we have plotted the molar ratios of Fe/Mg vs Si/Mg for all stars within the Hypatia Catalog (orange) that have abundances for the three elements -- and corresponding Mg measurements within the same data set. For a breakdown on how to convert stellar abundances to molar fraction, see Section \ref{chem}. We have overlaid the ratios of our sample in black, including the Sun as reference, per the values in Table \ref{ratios}. We see that the 10 closest stars are mostly centered within the plot, similar to the majority of the Hypatia stars, with a couple in the more extreme regions, namely HIP 99461 and HIP 73184. Similarly, we have plotted Si/Mg vs Al/Mg in Figure \ref{SiMg} (right), where the Hypatia stars are color-coded blue and the sample of 10 closest stars is black. Again, HIP 99461 and HIP 73184 are more extreme than the other stars in their molar ratios, in addition to HIP 12114. Finally, we show Si/Mg with respect to Ca/Mg in Figure \ref{SiMg} (bottom), with the Hypatia stars in light green. The sample of 10 stars show molar ratios that are clustered together as compared to the Hypatia stars. Both HIP 3765 and HIP 17378 have relatively low Ca/Mg while HIP 64394 is at the other extreme.  

\subsection{The Importance of the C/O Ratio}
The C/O ratio of a planetary disk is a primary control on the oxidation state of the condensates from which terrestrial planets are built. Above the C/O $\sim$ 0.8-1.0 threshold, carbon as graphite (SiC and TiC) becomes the dominant condensate at refractory temperatures \citep{bond_2010_aa}. Stars in these systems are likely to produce terrestrial planets dominated by similar reduced carbon species \citep{bond_2010_aa}. The high C/O systems are also unlikely to produce geodynamically active planets \citep{Unterborn14}, thus limiting degassing and any potential to be habitable. 

Refractory carbon may be present in disks of solar composition; \citet{Lodders03} notes that this will likely only be in small amounts, thus limiting the effects on final planet chemistry. \citet{Moriarty14} observed that the small amount of carbon may be as low as C/O $\sim$0.65 when dynamical effects within the disk are taken into account. Given the impact of carbon on planetary formation, our sample of 10 stars was chosen such that C/O $\ltaprx$ 0.8 with the exception of and HIP73184 and HIP12114 where either C, O, or both abundances were not available. Additionally, HIP108870 has C/O = 0.88, since we assume that the threshold for needed to take carbon chemistry into account within the disk as closer to 1.0. Future models and observations will likely help to constrain and narrow the C/O range of influence.

\section{Building Planets}\label{exoplex}
For those planets forming with C/O $\ltaprx$ 0.8, the dominant terrestrial planet-building condensates will be Fe and Mg- and Si-bearing silicates (e.g. forsterite Mg$_2$SiO$_4$), with comparatively minor portions of Al- and Ca- species (e.g. corundum, Al$_2$O$_3$) present as well. While the specifics of planet formation and accretion may cause a greater or lesser fraction of accretionary material to coalesce into planets, little fractionation of these major elements \textit{relative} to each other is expected. This is because in the case of Mg, Fe and Si, the elements have condensation temperatures within 20 K of each other for both solar and non-solar disk compositions \citep{Lodders03,Unterborn_2016}. When post-condensation dynamics are taken into account, the relative ratios (e.g. Si/Mg, Fe/Mg) are again found to change only on the order of 3 wt\% for our solar system \citep{bond_2010_aa,bond_2010_bb}  compared to solar abundances. Thus, while metallicity and other absolute abundances of the elements are potentially useful for understanding other aspects of exoplanets (e.g. the mass-metallicity relationship for Jovian exoplanets, \citealt{Thorngren_2016}), the major controls on terrestrial planet chemistry and mineralogy are the ratios of these 5 refractory elements: Mg, Al, Si, Ca, and Fe. 

The relationship between stellar composition and terrestrial planets is well grounded in our understanding of the chemistry and physics of planet formation. Recent work has shown the Sun's refractory composition is an acceptable proxy for the Earth's bulk composition in reproducing the Earth's mass, radius, and bulk composition to within 20\% \citep{Dorn15,Unterborn_2016}. The same is true for planets outside of our solar system with respect to their host star's refractory element abundances \citep{Thiabaud15, Dorn16, Dorn17}. These compositional contrasts offer hope given the the significant degeneracy in the inferred bulk interior structure and mineralogy of terrestrial exoplanets when mass and radius alone are adopted \citep{Dorn15,Dorn16,Dorn17}. 

The ExoPlex code iteratively solves for a planet's density, pressure, gravity and adiabatic temperature profiles that are consistent with the pressures derived from the mass within a sphere,
\begin{equation}
\label{MR1}
\frac{dm(r)}{dr}=4\pi r^{2}\rho(r)
\end{equation}
\noindent{the equation of hydrostatic equilibrium,}
\begin{equation}
\label{MR2}
\frac{dP(r)}{dr}=\frac{-Gm(r)\rho(r)}{r^2}
\end{equation}
\noindent{and the equation of state (EOS),}
\begin{equation}
\label{MR3}
P(r)=f(\rho(r),T(r))
\end{equation}
where $r$ is the radius, $m(r)$ is the mass within a shell of radius $r+dr$, $\rho$ is the density, $P$ is the pressure, $T$ is the temperature, and $G$ is the gravitational constant. The positions $r$ of the shells is then recalculated using the volume calculated from the new density at depth and shell mass. This process is then iterated until convergence, which we define as the change in density in every shell between iterations does not change by one part in 10$^{-6}$. We partition each modeled planet into a metal core composed of pure liquid-Fe and a rocky mantle. ExoPlex determines the mineralogy and density as determined by the EOS at each depth in the rocky mantle using the PerPlex Gibbs free energy minimizer package \citep{Conn09}.  We adopt the thermally dependent EOS formalisms of \citet{Stix05} for the mantle, \citet{Ander94} for the liquid Fe-core. While we neglect light elements in the core, the change to the mass-radius relation is only a few percent at Earth-like light element mass fractions \citep[$\geq 10\%$,][]{Unterborn_2016}.

\begin{figure*}
    \centering
    \includegraphics[width=\linewidth]{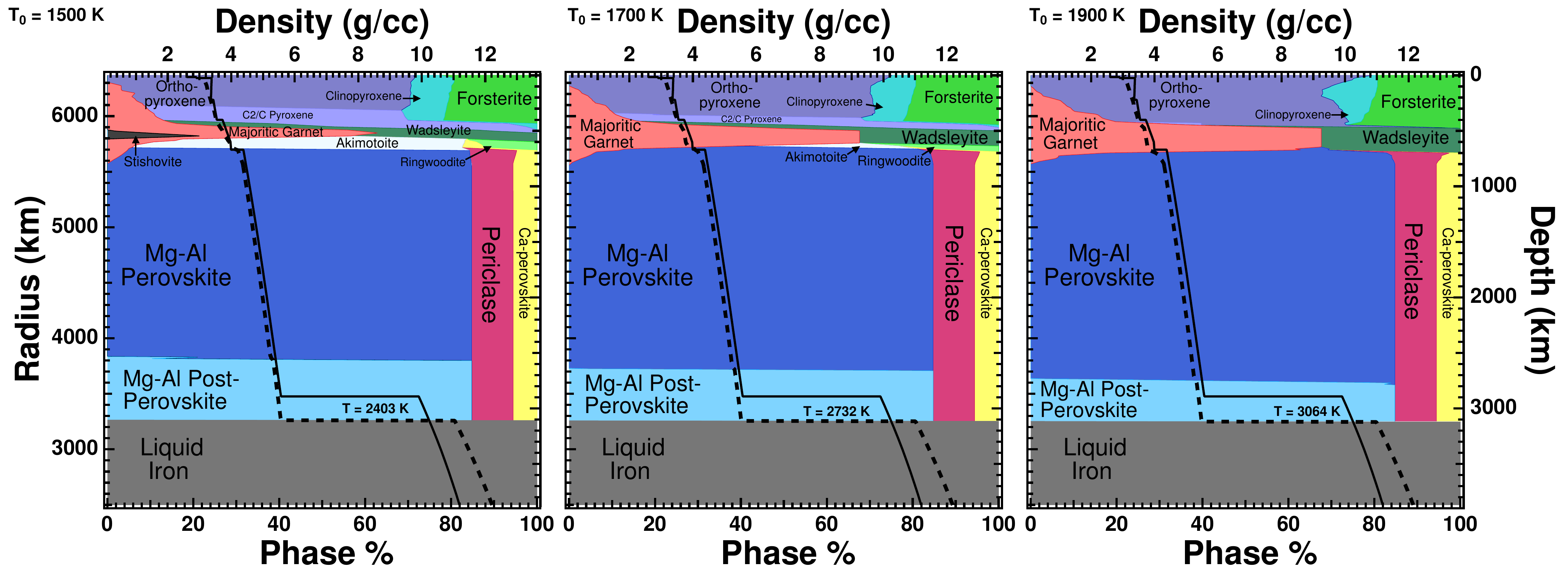}
    \caption{Phase diagrams for a modeled terrestrial planet of solar composition \citep{Lodders:2009p3091} for mantle adiabats of 1500 (left), 1700 (middle), and 1900 K (right). Density profiles are included for the modeled planet (black dashed) and the Preliminary Reference Earth Model \citep[PREM; ][]{PREM} is shown for comparison (black solid). Models were run assuming a planet of 1 Earth radius.}
    \label{fig:Sun}
\end{figure*}

For our sample of 10 stars, initial ratios from Table \ref{ratios} were adopted and model mineralogies and density profiles were calculated for fictive terrestrial planets orbiting the stars, assuming to be 1 Earth-radius. Models were run using the ExoPlex planet-building code \citep{Unte17c,Lore17} along self-consistent adiabatic temperature profiles of 1500, 1700 and 1900 K, which cover a range of cold, hot, and ``Earth-like" geotherms. For simplicity, all Fe is assumed to remain in the core and thus represents a planet with an oxidation state at or below any oxidized iron redox buffers (e.g. iron-w\"ustite/quartz-iron-fayelite). We show in Figure \ref{fig:Sun} the modeled mineralogies and density profiles for a planet of solar composition \citep{Lodders:2009p3091}. Overlaid as a solid black line is the density profile of the Preliminary Reference Earth Model \citep[PREM; ][]{PREM}. While the PREM incorrectly predicts the size of the core (likely due to the lack of incorporating light elements into the metal), the density profiles of the ``solar" planet and Earth are remarkably similar. 

Similar models for Venus would elucidate the validity of our star/planet model, since Venus makes up nearly 40\% of the total mass of terrestrial planets in the solar system. The bulk composition of Venus or even its core size (a key constraint in mass-radius models) is not known. Mars and Mercury, on the other hand, have modest constraints on their bulk composition, however, given their small sizes (0.10 and 0.06 Earth masses, respectively, or $<$10\% of all terrestrial planet masses in the Solar System), low-pressure silicate phases will dominate the bulk of the silicate portion of the planet. These minerals will be more indicative of the Earth's upper mantle, where changes in oxygen fugacity and minor element content will have a larger effect on the phase equilibria and behavior of the minerals. The threshold from upper-mantle-dominated planets occurs at roughly 0.2 Earth masses, and we would stress the lack of utility in applying star/planet models to any future exoplanet discoveries below this mass. Further details will be provided in Paper II, \citet{UnterbornHinkel}.

The compositional space for the sample of 10 stars are similar and not wholly unlike the Earth, with varying proportions of the dominant terrestrial planet-building minerals: olivine, pyroxene, Mg-Al-perovskite and periclase. Temperature has a minor effect, mostly changing mineralogies in the transition zone, from akimotoite-dominated for colder planets and wadsleyite-dominated for planets with potential temperatures greater than the modern Earth's. While these models do not include the effects of Fe incorporation into these minerals, it should be noted that Fe partitioning into each of these species is possible. Geochemical and geophysical consequences of the abundance variations are explored more generally in forthcoming work \citep{UnterbornHinkel}. Of note, though, are the similarities between the mineralogies and structures for planets around these 10 stars. As shown in Tables \ref{abunds1} and \ref{abunds2}, the \textit{bulk abundances} vary to a significant degree between the stars, especially for the thick-disk star HIP 57939 which has [X/H] abundances consistently below -1.0 dex. However, the \textit{molar fractions} of these elements (see Section \ref{chem}) are markedly similar, per Table \ref{molarerr} and Figure \ref{SiMg}. 

ExoPlex utilizes the spreads associated with the stellar abundances from the Hypatia Catalog, per error propagation of the two abundances used to calculate that molar fraction; for example the error for Si/Mg is determined from the spread associated with [Si/H] and [Mg/H]. The large spreads in the abundances create significant errors in the molar fractions, as shown in Table \ref{molarerr}. As a result, and as discussed more extensively in Paper II \citet{UnterbornHinkel}, the compositions for the 10 planets are markedly alike to within the error or spread.

\section{Determining Distinct Planetary Mineralogies}\label{chem}

We determined the errors ($\sigma$) for the molar fractions, shown in Table \ref{ratios}, by propagating the spread in the stellar abundances (per Tables \ref{abunds1} and \ref{abunds2}) for the anti-logarithm, base-10. In other words, the molar fraction X/Y is determined by the stellar abundances X = [X/H] and Y = [Y/H], such that
\begin{equation}
X/Y = 10^{(X+A-Y-B)}\,\,,
\end{equation}
where A is the solar composition of X and B is the solar composition of Y. We calculate the error on the molar fractions, S$_{X/Y}$, per:
\begin{equation}\label{molarerr}
S_{X/Y} = 2.303 * \sqrt{S_{X}^2 + S_{Y}^2}\,,
\end{equation}
where S$_{X}$ and S$_{Y}$ are the spreads or errors on [X/H] and [Y/H], respectively, as given in Tables \ref{abunds1} and \ref{abunds2}. For our purposes, we assumed that there was no error on the solar stellar abundances. Additionally, we utilized the spread in the Mg values associated with each respective element (see Section \ref{sample-abunds}). In this way we calculated the $\sigma$ errors shown in Table \ref{ratios}. Looking at all of the ``extreme" molar fractions from Figure \ref{SiMg}, it's clear that the errors overlap for all of the stars in our sample. 

\begin{figure*}
\centering 
\includegraphics[height=6.5cm]{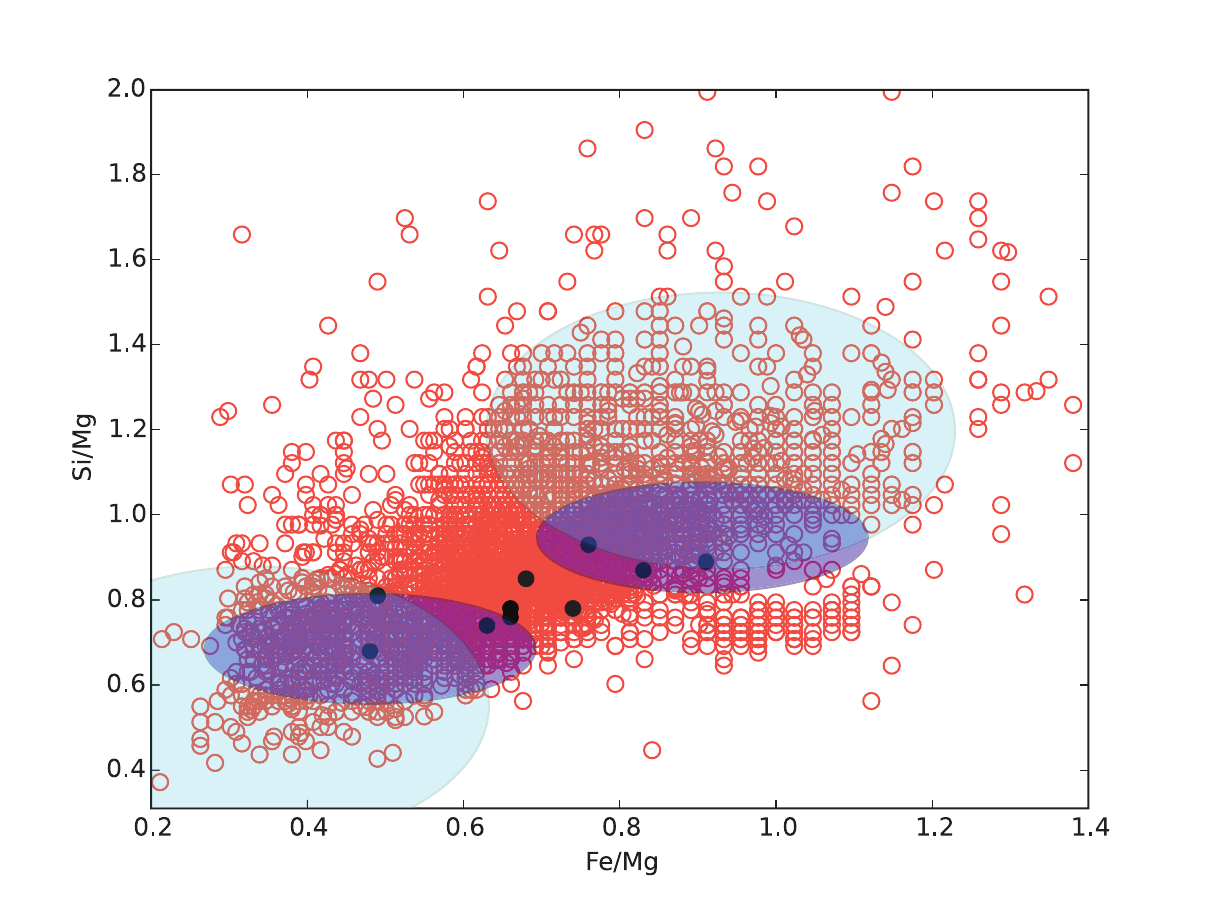}\quad
\includegraphics[height=6.5cm]{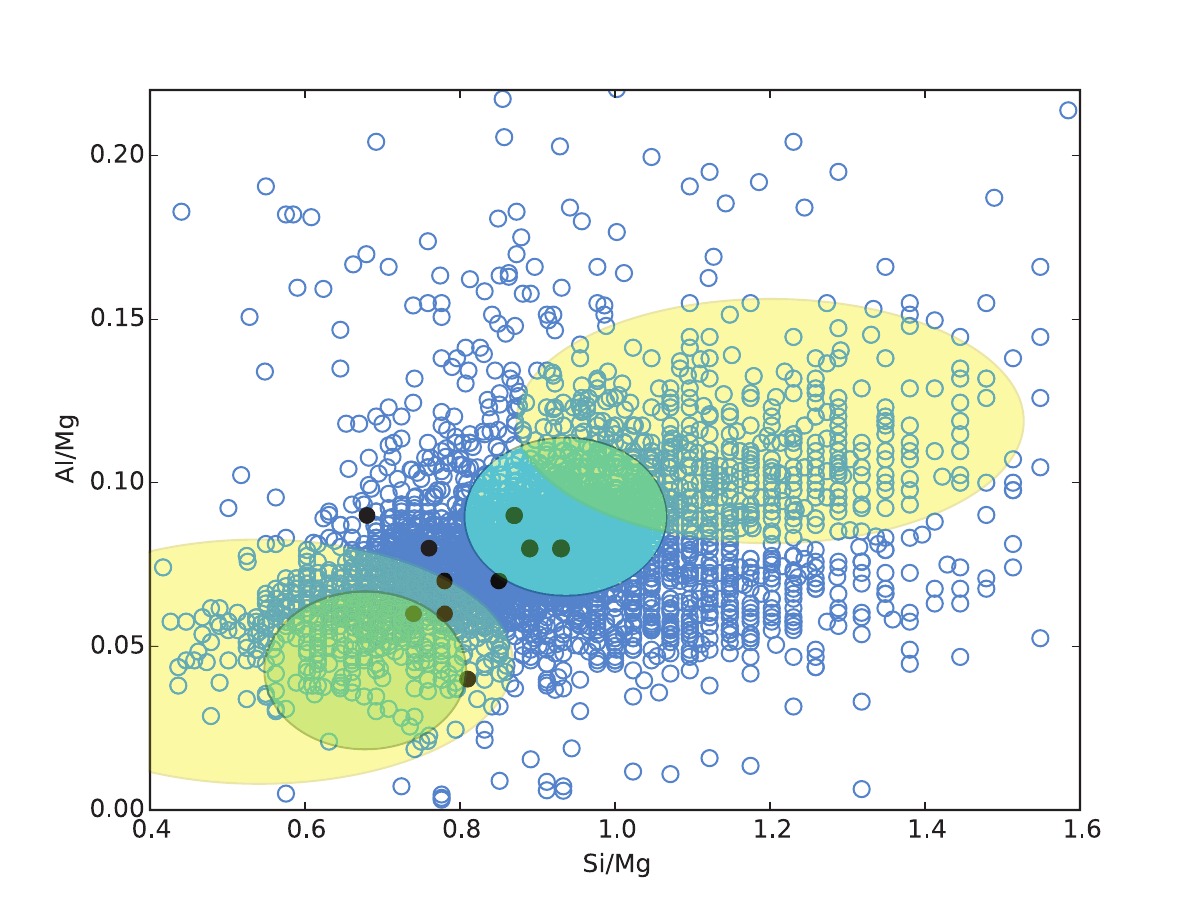}\par\medskip
\includegraphics[height=6.5cm]{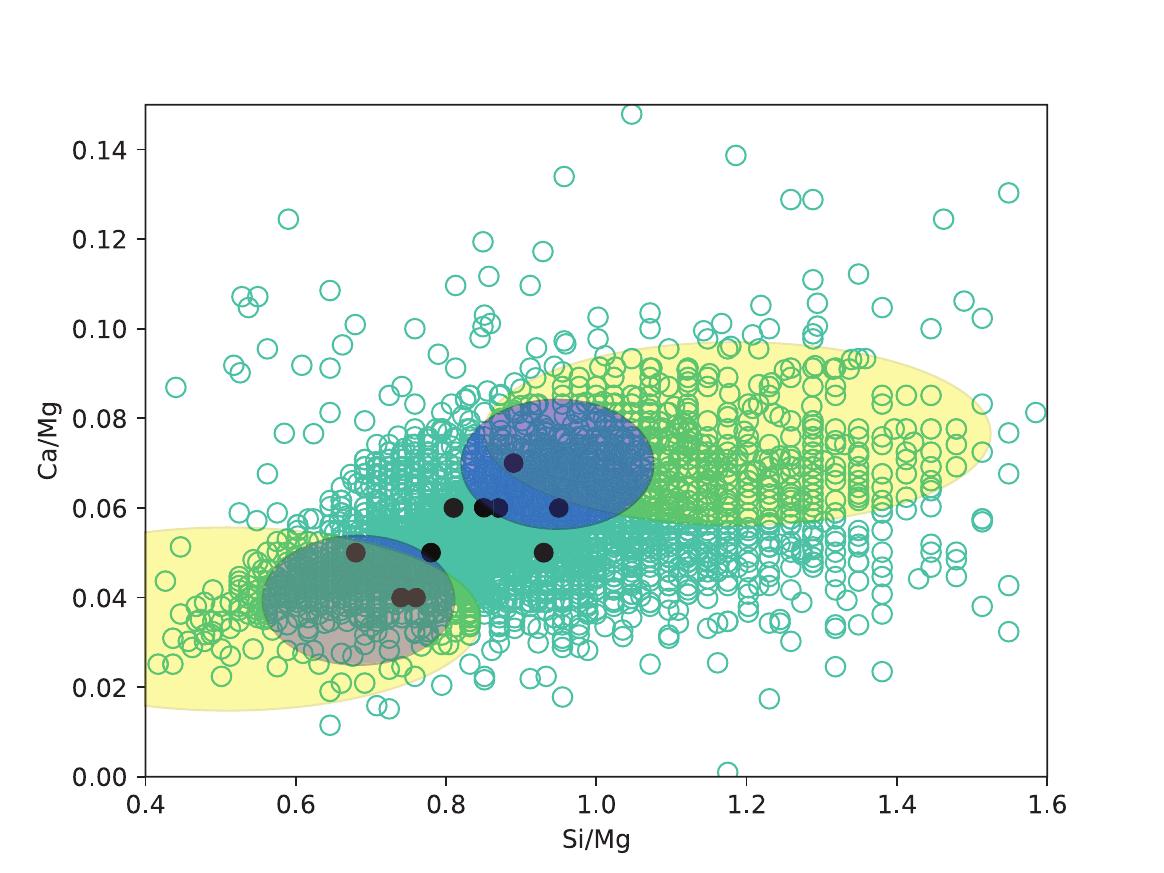}

\caption{Reproduction of the subplots in Figure \ref{SiMg} without the Sun, with error ellipses that span half the range of the sample in both the x- and y-directions. Left: the dark blue ellipses show the extremes for the sample of 10 stars (black dots) studied here while the light blue ellipses are with respect to the full sample in orange from the Hypatia Catalog. Right: similar as the left, where the light green ellipses are for the 10 stars and the yellow are for the Hypatia stars. Bottom: similar to the other two, where the light blue ellipses are for the 10 stars and the yellow are for the Hypatia stars. See text for details.}\label{ellipses}
\end{figure*}

However, while the spreads in the [X/H] stellar abundances are large, we would like to know what the spreads or ultimate error within the stellar abundances must roughly be {\bf in order to distinguish two unique planetary populations}. Therefore, we compute the general precision that stellar abundance errors or spreads need to achieve for the molar ratio errors to meet half way between the ``extreme" stars in our sample, not including the Sun. In this way, we can identify what stellar information is needed for the planetary models in our sample to be ``different" to within error. Per Table \ref{ratios}, the extreme Si/Mg ratios are found in HIP 99240 and HIP 23311; for Fe/Mg we look at HIP 64394 and HIP 23311; for Al/Mg we examine HIP 23311/1088870 and HIP 57939; and finally for Ca/Mg we look at HIP 64394 and HIP 3765/17378. For Si/Mg the max distance between the extreme stars is 0.25 so half of that is 0.125, which we will call $\sigma_{max}$. For Fe/Mg, half of the max $\sigma_{max}$ is 0.215, for Al/Mg it's 0.025, and for Ca/Mg it is 0.015. 

We give a pictorial diagram of the $\sigma_{max}$ errors for our sample in Figure \ref{ellipses}. On the left, we reproduce Figure \ref{SiMg} (left) showing the sample of 10 stars as black dots and the Hypatia stars in orange. We have overlaid dark blue ellipses to show the extreme molar ratios between the stars in our subsample. In the case of the lower-left dark blue ellipse, the extreme low in both Si/Mg and Fe/Mg is HIP 23311. However, for the upper-right dark blue ellipse, there isn't one single star with the highest Si/Mg and Fe/Mg for this sample, so we put the center of the ellipse where it would be. The ellipses have a width and height of extreme high-to-low range, or  2 $\times$ $\sigma_{max}$ values, for the respective molar ratios, such that the edges of the ellipses only just touch in the x- and y-directions. By separately analyzing the molar ratios in the x- and y-directions, the ellipses are in each of the two directions individually, but they did not physically touch in the 2D-plane. If the dark blue ellipses touched in the 2D plane, there would be an overlap in Si/Mg and Fe/Mg between the two populations, which would produce a degeneracy in the planetary compositions. By removing any overlap, we have created distinct populations of stars that are separate to within error in both the x- and y-directions. 

To the right of Figure \ref{ellipses}, we have recreated Figure \ref{SiMg} (left) where we have again added green ellipses at the extreme molar ratios within the sample of 10. Since the same star does not have both the highest and lowest molar ratios of Al/Mg and Si/Mg, we have placed the center of the ellipses where the extremes occur (see Table \ref{ratios}). Again, the entire range of the $\sigma_{max}$ values in the x- and y-directions are encompassed by the width and height, respectively, of the ellipses. Finally, on the bottom of Figure \ref{ellipses}, we have plotted Ca/Mg with respect to Si/Mg (as seen in Figure \ref{SiMg}, bottom), where the center of blue ellipses represent the extremes for both molar fractions. 

We take these error maximum values as S$_{X/Y}$ = $\sigma_{max}$, or half the width and height of the ellipses. Then, assuming that the spread or error in [X/H] is the same as [Y/H], or S$_x$ = S$_y$, to make the calculations simpler, we can simplify Eq. \ref{molarerr} to be:
\begin{equation}
\sigma_{max} > 2.303 * \sqrt{2 * S_x^2}\,\,.
\end{equation}
From here, we solve for S$_x$, which results in: 
\begin{equation}
\sqrt{0.5*\bigg( \frac{\sigma_{max}}{2.303} \bigg) ^2} > S_x\, .
\end{equation}\label{sigmamax}
We now have an equation that will calculate the spread or error in the stellar abundances that will in turn produce an error in the molar ratio that spans half the range (or less) of our sample. 

For the Si/Mg molar ratio, $\sigma_{max}$ = 0.25 so the [Si/H] and [Mg/H] stellar abundances need a precision of 0.01 dex or better, per Eq. \ref{sigmamax}. Similarly, for the Fe/Mg ratio, [Fe/H] and [Mg/H] need a 0.02 dex precision (Figure \ref{SiMg}, left). The [Al/H] and [Mg/H] abundance ratios need 0.002 dex for Al/Mg (Figure \ref{SiMg}, right) while [Ca/H] and [Mg/H] need 0.001 dex for Ca/Mg. In other words, the stellar abundances have to be precise on the order of 0.001--0.02 dex in order to distinguish two distinct populations within the molar ratios. While very difficult, there are a number of groups who have reported that precision for large groups of stars, see Section \ref{highprecision}. With precise stellar abundances below these limits, we can define two unique populations of stars (as seen in Figure \ref{ellipses}). Namely, if terrestrial planets were discovered in orbit, our model predicts the planetary structures and mineralogies would be distinctly different to within error of both the stellar abundances and molar ratios. The third population of planetary systems, namely those that are not included within the ellipses, would have degenerate compositions that could not be indicated as either having low molar ratios of Fe/Mg vs Si/Mg vs Al/Mg vs Ca/Mg or high molar ratios.

\begin{figure*}
    \centering
    \includegraphics[height=6.5cm]{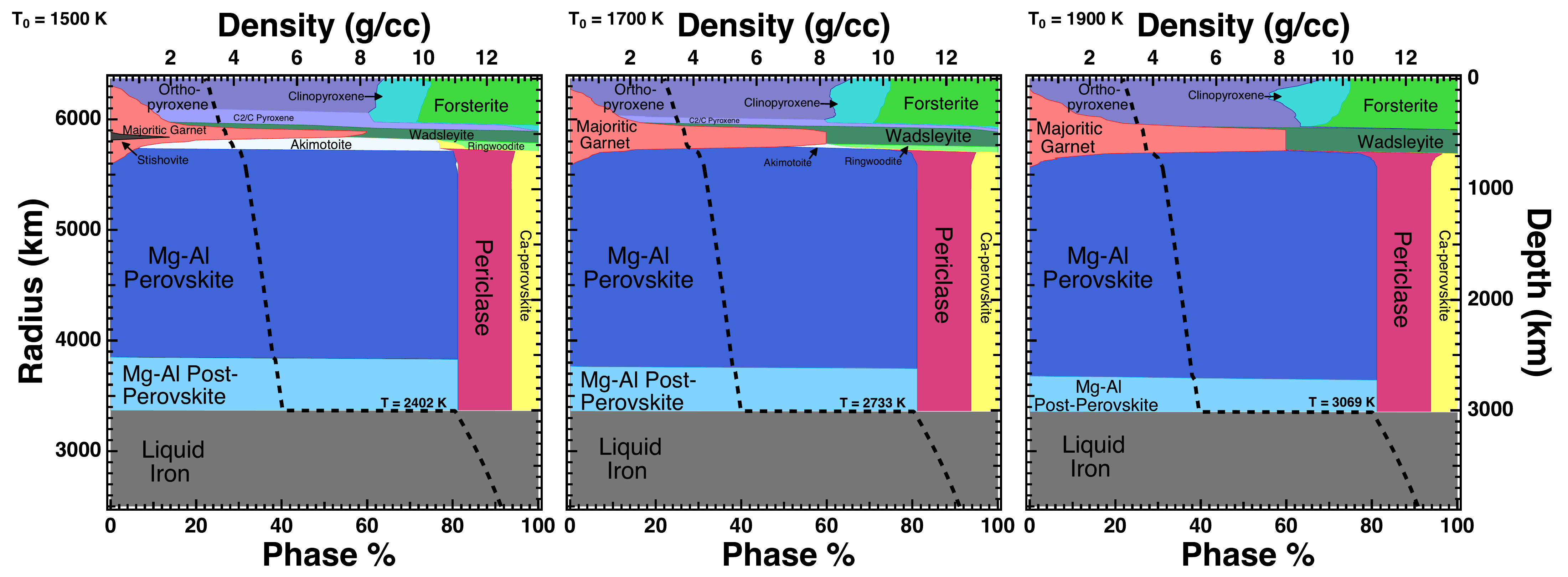}\quad
    \includegraphics[height=6.5cm]{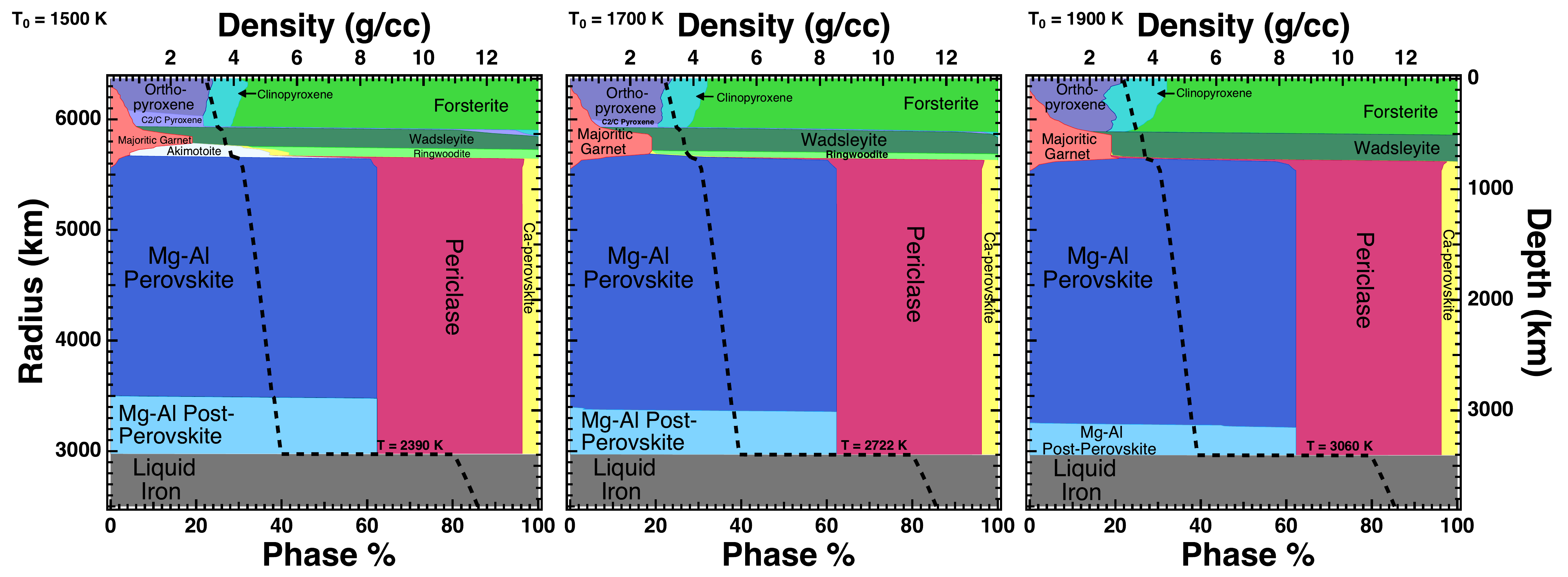}

    \caption{Planetary minerology phase diagrams, similar to Figure \ref{fig:Sun}, representing the extreme high molar fractions (top) and extreme low molar fractions (bottom) as outlined in the text and represented as ellipses in Figure \ref{ellipses}. The most notable differences are the sizes in core and variation in Si/Mg which affects many aspects of the structure. See text for details.}
    \label{highlowmodels}
\end{figure*}

Using ExoPlex (see Section \ref{exoplex}), we have modeled the planetary compositions of the extreme high and extreme low molar fractions, taking into account the uncertainty required in our analysis to make the two distinct -- see Figure \ref{highlowmodels}.  As expected, the high Fe/Mg population (top) has a larger core than the low Fe/Mg planet (bottom). In addition to this major structural difference, the lower Si/Mg sample contains a larger fraction of both olivine (in the upper mantle) and periclase (in the lower mantle) than the greater Si/Mg population. Furthermore, the transition zone mineralogies are significantly different, with the low Si/Mg sample containing little comparative garnet, thus increasing the overall water storage capacity of this key mantle reservoir. Temperature, too, affects the mineralogy of the transition zone as well. With those planets run along a ``hot" adiabat (1900 K mantle potential temperature), displaying no ringwoodite, whereas the ``cold" adiabat stabilizes stishovite (a high pressure polymorph of SiO$_2$) and akimotoite at the expense of ringwoodite. While these are only those indicative from a first-order, equilibrium mineralogy model, these compositional changes can affect the dynamical state and rate of geochemical cycling on these planets. 

\subsection{Planet Mineralogies within Hypatia}\label{hypmins}
In this paper, we have chosen to look a small sample of nearby ($<$ 10 pc) stars. However, it is clear that those stars encompass only a small fraction of the molar ratio space, namely Fe/Mg vs Si/Mg vs Al/Mg vs Ca/Mg, as shown in Figure \ref{ellipses}. So as not to say that abundance ratios of refractory elements do not expand beyond this range, we expand our exploration to look at the full sample of the Hypatia stars, per Figure \ref{ellipses}. In this way, we can analyze how the errors in stellar abundances propagate to molar fraction and, ultimately, how that impacts planetary structure and mineralogy for $\sim$6000 stars in the solar neighborhood. We look at the extreme values for the molar ratios that have a sufficient density within the plots and place the centers of the light blue (left), yellow (right), and yellow (bottom) ellipses at those locations. For the Si/Mg vs Fe/Mg plot (left), that occurs at (0.33, 0.55) and (0.92, 1.20) while the Al/Mg vs Si/Mg plot (right) has ellipses centered at (0.55, 0.04) and (1.20, 0.12). In the bottom plot, Ca/Mg vs Si/Mg, the ellipses are centered at (0.55, 0.035) and (1.20, 0.075). To determine two populations of stars that are different to within error of the molar ratios, we define the major- and minor-axes of the ellipses as the total extreme, high-to-low range for the respective molar ratio such that the two ellipses just touch, but do not overlap, in each Cartesian direction. 

Half of the extreme range in the molar ratios is $\sigma_{max}$, which is 0.325 for Si/Mg, 0.295 for Fe/Mg, 0.04 for Al/Mg, and 0.02 for Ca/Mg. Going through a similar calculation as before using Eq. \ref{sigmamax}, we find that both Si/Mg and Fe/Mg molar ratios require spreads or errors in stellar abundances [Si/H], [Mg/H], and [Fe/H] that is at or below 0.03 dex. The Al/Mg ratio finds that [Al/H] and [Mg/H] need a precision of 0.004 dex. Finally, the Ca/Mg ratio requires [Ca/H] and [Mg/H] need a precision that is 0.002 dex. In other words, in order to calculate planetary structures and mineralogies that are different, to within error of the molar ratios, the respective stellar abundances need to be precise to 0.03 dex or below. Given that the Hypatia sample of stars spans nearly twice the range in molar ratios as compared to the sample of 10 stars, the precision in stellar abundances is not greatly expanded. We will go into further detail on how high precision stellar abundances may be obtained in Section \ref{highprecision}.

\subsection{Assumptions in the Planet Mineralogies} 
The purpose of this walk-through is to obtain a general idea for the precision needed in stellar abundances to produce (only two) distinct populations of terrestrial planets. However, we made some assumptions along the way, for simplicity's sake, that have minor effects on our calculations. For example, we assumed that S$_X$ = S$_Y$ or that the error or spread in [X/H] was equal to [Y/H]. In general, the error reported for stellar abundances of [Mg/H], [Al/H], [Si/H], and [Ca/H] are 0.07 dex, 0.06 dex, 0.05 dex, and 0.06 dex, respectively \citep{Hinkel14}. Looking at the some of the higher-precision abundances, \citet{Nissen15} reported errors of 0.009 dex, 0.005 dex, 0.005 dex, and 0.006 dex, respectively, while \citet{Spina16a} had errors of 0.014 dex, 0.012 dex, 0.007 dex, and 0.013 dex. In total, it appears as though our assumption that S$_X$ = S$_Y$ isn't radically far from the truth, although the variation could contribute somewhat to the needed stellar abundance precision. 

When determining stellar abundances, not all datasets are on the same baseline with respect to the solar normalization (see Eq. \ref{dex}). Namely, different groups chose one of a variety of solar measurements, for example \citet{Anders:1989p3165, Grevesse:1998p3102, Asplund:2009p3251, Lodders:2009p3091}, or simply measured their own either directly or as reflected light at the time of observations. To date, there are 45 individual solar normalizations taken into consideration within the Hypatia Catalog. The assortment of solar normalizations introduces an intrinsic scatter when comparing stellar abundances. For example, in Hypatia, the range in the absolute abundance of Mg in the Sun is 0.17 dex while absolute Fe has a range of 0.26 dex. As a result, these solar normalization disparities create a variation of $\sim$0.1 dex in the molar ratios (see Table \ref{ratios}). Within the Hypatia Catalog, we sought to correct the baseline differential by renormalizing all stellar abundances to the same solar scale, namely \citet{Lodders:2009p3091}. However, for the analysis in this paper, we assumed that there were no errors on the solar composition. Looking at \citet{Lodders:2009p3091}, they report absolute errors of Mg = 0.06 dex, Al = 0.07 dex, Si = 0.01 dex, and Ca = 0.02 dex for the Sun while \citet{Asplund:2009p3251} lists Mg = 0.04 dex, Al = 0.03 dex, Si = 0.03 dex, and Ca = 0.04 dex. These solar abundance errors can contribute an additional $\sim$0.007 dex precision needed in the stellar abundances to obtain two separate populations of terrestrial planets via the molar fractions.

Finally, we have used a specific subsample of 10 nearby stars in order to demonstrate our intention of determining two unique populations of planetary mineralogies. While we have expanded that demonstration to encompass the entire Hypatia Catalog in Section \ref{hypmins}, our choice of location for the ellipses was approximate and not overly rigorous. Therefore, we urge caution that the required abundance precisions quoted here are to be used a general criterion, such that they are tailored to meet the specifics of any future study.

\subsection{High Precision Stellar Abundances}\label{highprecision}

Through the works of \citet{Torres12, Smiljanic14, Hinkel14}, it has become apparent that stellar abundance techniques are discrepant. In \citet{Hinkel16}, an international team of stellar abundance groups came together to uncover the underlying reason as to why stellar abundance measurement techniques varied. The study supplied six groups with the same stellar spectra and tested the effects of standardizing the stellar atmospheric parameters (namely, \teff and \lg), the element line lists used to measure the abundances, and both in tandem. While some standardization helped somewhat, the kind of standardization and the extent varied between elements as well as methodologies. Ultimately, the experiment was not able to completely reduce the spread in abundances between groups. 

In order to determine whether a planet's interior structure falls into a low molar ratio, high molar ratio, or neither/both category, we found here that stellar abundances need to be nearly an order of magnitude more precise than current, individual measurement techniques allow. Additionally, those measurement techniques need to be corroborated such that the range or spread of abundance measurements for the same element in the same star approaches zero. By applying the spread as the associated error in this study, we have utilized this important metric to illuminate how well-measured stellar abundances truly are.

With that in mind, the abundance precision levels needed to calculate at least two independent populations of planetary interior structures are not impossible. A number of individual groups have been able to obtain stellar abundances with precisions in the thousandths of a dex, as mentioned earlier. For example, \citet{Ramirez14a} employed a differential approach that determined stellar abundances with respect to stars other than the Sun, namely HIP 74500 and HIP 14954, that were more similar in stellar properties with respect to the rest of their sample. By using this technique, they were able to remove systematic errors typically associated with temperature or metallicity. In a similar vein, \citet{Nissen15, Nissen16} measured the abundances in only those stars from the \citet{Sousa08} sample that were solar twins, or stars that were within $\pm$100 K in \teff, $\pm$0.15 in \lg, and $\pm$0.10 dex in [Fe/H] as compared to the Sun. \citet{Spina16a} chose a set of 14 solar twins based predominantly on their color in both the optical and infrared spectrum as compared to the Sun. \citet{Adibekyan16} worked with a set of 40 stars that had ages close to that of the Sun. Additionally, while they limited \teff and \lg to be solar-like, they noted that the variation in their stellar parameters was likely the reason that they were not able to achieve the precision determined by other groups. 

Finally, the implementation of local thermodynamic equilibrium (LTE) vs. non-LTE within stellar atmospheric models has been found to yield dramatically different results for a number of elements \citep[i.e.][]{Gehren06}. In some cases, the effects of NLTE on certain elements is small, for example in \citet{Bensby14, Luck17}. However, NLTE seems to be particularly important for stars with a low metallicity, or [Fe/H] $<$ -1.0, where the the stellar models deviate from solar \citep{Zhao16}. On the other hand, a line-by-line differential method to determine the stellar abundances, as implemented by many of the above groups, is useful for cancelling out the difference between LTE and NLTE. 

Overall, in order to achieve high precision, the sensitivity of the stellar atmospheric models must be accounted for, such that the similar stars are compared to one another per the abundance ratios. Additionally, extremely high resolution spectra is required. While this approach severely limits the number of stars for which planetary interiors can be distinguished, it does highlight a path forward.

\section{Physical Properties When Defining Habitable Planets}\label{physical}

The chemical and physical properties of a star need to be considered in tandem. While we have talked about the stellar abundances, we now consider the physical characteristics of the star that are important for habitability. For example, the measurement of stellar activity (via Ca II H and K lines defined as the $R^{'}_{HK}$ index) indicates the strength of stellar magnetic field, which is directly responsible for the structure of the corona and propagation of solar winds and flares. Therefore, it is important for habitability that a star have relatively low stellar activity. In this section we analyze the properties of the stellar systems as it pertains to the habitability of potential terrestrial planets.
Additionally, we have provided both the optimistic and conservative HZ radii \citep{Kopp13,Kopp14} for all 10 stars in Table \ref{physHZ}, indicating where liquid water could be expected on the surface of a planet. We have also determined an estimate of the periods and radial velocity measurements expected for potential Earth-like planets orbiting the ten stars at both the conservative and optimistic HZ radii, per the equations in \citet{Kane07}. Since we do not have stellar masses for our sample, we calculate an approximation using a main-sequence mass-temperature relationship: $(M/M{_\odot})^{2.5} = (T/T{_\odot})^4$, the values of which are listed in Table \ref{physHZ}. In this way, we hope to provide a reference for the RV precision required to detect potential planets in the context of upcoming space missions like PLATO, which will be dedicated to searching for these kind of planets.

\subsubsection{HIP 108870 -- $\epsilon$ Ind}

As shown in Table \ref{phys}, HIP 108870 is the closest star in our sample. \citet{Endl02} used the radial velocity (RV) technique, or the ``wobble" method which gauges the motion of a star's center of mass due to a companion, and found that the system had a low-amplitude linear trend. This behavior was determined to be a brown-dwarf binary system separated by 2.65 AU at a distance of 1459 AU \citep{Volk03, McCaughrean04, Scholz03}. Because of the large separation between the main sequence star and the brown-dwarf binary (note: Pluto is $\sim$40 AU from the Sun), we found that the companion had little impact on the primary star and therefore we kept this target within the sample. \citet{Janson09, Zechmeister13} both came to a similar conclusion when looking for nearby, Jupiter-mass planets, such that the binary-dwarfs could only account for an acceleration of 0.009 m/s/yr, an extremely small effect. Long-term trends were found in both studies in the RV, namely 2.4 m/s/yr \citep{Zechmeister13}, and with respect to stellar activity per $\log$ R$^{'}_{HK}$. However, while the trends could be explained by massive planetary companion, there has been no detection of a giant planet. Additionally, the authors noted that the trends in RV and $\log$ R$^{'}_{HK}$ could be coincidental. Due to the high precision RVs and imaging via HST/NICMOS and VLT/NACO \citep{Geissler07,Janson09}, it is safe to conclude that HIP 108870 is not an active star. 
Despite the presence of an extremely wide brown-dwarf binary companion, the literature reports that HIP 108870 is a relatively quiet, inactive star. And while there appear to be indirect signs of a possible orbiting planet, the search for one has not been successful. It is clear from the multiple previous observations that HIP 108870 is an excellent star for hosting a planet that has a good chance of being habitable from a physical perspective.

\subsubsection{HIP 96100 -- $\sigma$ Dra}

HIP 96100 has an estimated radius
of R = 0.778 $\pm$ 0.008 R$_{\odot}$ and age of $\sim$5-10 Gyr based on multiple stellar isochrones per \citet[][and references therein]{Boyajian08}. HIP 96100 is an RV constant star such that it was used to track the zero-point drift of HIRES on the 10.2 Keck telescope in \citet{Courcol15}. Therefore it is likely there are no giant exoplanets in the system. 
Ultimately, the constant nature of the stellar activity means that HIP 96100 has all the physical qualities to host a habitable, terrestrial planet. 

\subsubsection{HIP 99240 -- $\delta$ Pav}
HIP 99240 is relatively stable in the RV, ruling out any massive secondary companions either close to the star or in a wide orbit \citep{Wielen99}. Velocity oscillations via asteroseismology were found to be centered on 2.3 mHz, with peak amplitudes similar to the Sun \citep{Kjeldsen05}. Additionally, with respect to potential planet formation, there does not appear to be a cold dust disk around HIP 99240 \citep{Eiroa10}. In other words, if a planet is orbiting HIP 99240, it is likely a smaller, terrestrial planet. The physical similarity of HIP 99240 to the Sun means that it would be a good location for a habitable, Earth-sized planet. 

\subsubsection{HIP 3765 -- HD 4628}

HIP 3765 shows low levels of stellar activity per the Ca II H and K emission \citep{Mathioudakis94, Affer05}. The stellar age was determined to be 3 $\pm$ 1.5 Gyr based on theoretical isochrones in \citet{Affer05}, although we note that the [Fe/H] abundance value they used in that determination, namely -0.27 dex, is slightly lower than the value we report here in Table \ref{abunds2}: [Fe/H] = -0.20. The consistency of the stellar activity means such a planet around HIP 96100 would be physically stable in terms of habitability. 

\subsubsection{HIP 2021 -- $\beta$ Hyi}

HIP 2021 has been studied via astroseismology in order to 1) determine its mass and radius \citep{North07} as well as 2) measure the effect of stellar oscillation and granulation to minimize planetary detection limitations \citep{Dumusque11}. While a number of studies have analyzed the RV data of this target \cite[e.g.][]{Endl02}, that data was compiled by \citet{Zechmeister13} who found no obvious trend in the data. The lack of an RV trend may be due to a discrepancy in the RV data or perhaps a correlation with the star's magnetic cycle. Despite the low activity of the star and techniques to increase planet detectability, no giant planets have been discovered around HIP 2021. 

\subsubsection{HIP 7981 -- 107 Psc}
HIP 7981 is a chromospherically active star, which has an observed short-term pattern of starspots \citep{Messina99} and cyclical luminosity variations that can be multiple times that of the Sun \citep{Radick01}. From our literature search, it does not appear as though this star has been directly or pointedly observed, especially with respect to planetary surveys, in the last few decades. 

\subsubsection{HIP 23311 -- HD 32147}
HIP 23311 is a member of the HR 1614 moving group \citep{Eggen78, Eggen92}, namely an association of stars from the same stellar birth cloud, with similar galactic velocity, that was later disrupted by differential galactic rotation to form an elongated ``tube" \citep{Antipova2015}. While studying the chemical abundances for stars in HR 1614 group, \citet{Antipova2015} found that, unlike the other members that had [Fe/H] $\sim$ 0.2-0.3 dex, HIP 23311 was anomalous. Namely, the atmospheric parameters \lg and \teff were too low to be consistent with a dwarf star. Additionally, HIP 23311 had [Fe/H] = -0.14 dex while being enriched in Na, Mg, Al, and Si as compared to other members of the moving group. The authors concluded, in a separate study \citep{Antipova2016}, that the dwarf star exhibits solar-like activity in the form cool, dark, often large star spots. These star spots affected the equivalent widths of the spectral lines, which resulted in the anomalous and incorrect abundance measurements. The large star spots and corresponding high stellar activity, which can manifest in intense UV and X-ray radiation, could impose varying, perhaps unpredictable, physical conditions on an orbiting terrestrial planet \citep{Garraffo17}. 

\subsubsection{HIP 17378 -- $\delta$ Eri }

HIP 17378 has been classified as a variable of RS CVn type\footnote{http://simbad.u-strasbg.fr/simbad/sim-fbasic}, which are defined as close binary systems that have high chromospheric activity and star spots. However, there has been no observation of any stellar activity, including the X-ray, photometric, or emission from the H and K lines \citep[][and references therein]{Thevenin05}. Additionally, there is no RV variation or any considerable photometric change. Therefore, \citet{Thevenin05} pointedly searched for a companion star using the VLTI and found there was no companion to within $\pm$2\% of the luminosity of HIP 17378, or L/L$_{odot}$ = 3.19 $\pm$ 0.06. The authors conclude that the RS CVn classification is doubtful. The first three oscillation modes in HIP 17378 have also been observed, namely ($l$, $m$) = (1, −1), (2, 1), and (1, 0/-1), where the latter $m$-mode could not be disentangled \citep{Hekker10}. Their work implied that the non-radial modes were the dominant frequency -- a widely debated topic in red-giant stars at that time. Additionally, while the classification of RS CVn means that an orbiting planet would receive a lot of high energy radiation, the lack of any observed stellar activity is encouraging for habitability.

\subsubsection{HIP 57939 -- HD 103095 }
HIP 57939 originated from the halo of the Milky Way, making it the closest halo star to the Sun. It has also been chosen as a benchmark star for the Gaia mission \citep{Jofre14}. It is difficult to measure not only the \teff of metal-poor stars, but also to match observations with models of stellar structure and evolution and, therefore, to determine the fundamental properties of the star \citep{Creevey12}. Those stellar parameters found by \citet{Soubiran16} can be found in Table \ref{phys}. Age estimations for this star vary from 5.261 $\pm$ 4.089 Gyr to 10.19 $\pm$ 1.58 Gyr as laid out by \citet[][and references therein]{Mishenina17} to 12.0 $\pm$ 0.2$^{+1.8}_{-2.2}$ Gyr \citep{Creevey12}. 
Since little is known about the physical properties of the HIP 57939 star, it is unclear how it would influence an orbiting exoplanet.

\subsubsection{HIP 64394 -- $\beta$ Com }

The magnetic field of solar-like HIP 64394 was measured by \citet{Gray96, Plachinda99}. The latter used the S index per the H and K lines over a baseline of 5 years and found a $\delta$ S $\sim$ 0.025 dip in activity, the largest seen since 1966. Additionally, there was an photometric dip $\approx$ 5 mmag  and a temperature variation of $\approx$ 30 K during that same epoch as the magnetic activity \citep{Plachinda99}. The ultimate implication is that the magnetic change drove the variation in the $b$ and $y$ photometry and temperature. The physical properties and stellar activity of HIP 64394 are variable enough that it might detrimentally impact the habitability of a orbiting planet.

\section{Conclusion}\label{conc}

Defining planetary habitability and Earth-likedness is a difficult task, one that is compounded by the fact that there is only a single data point for reference, namely, the Earth. One of the most prevalent definitions of the habitable zone assumes ``Earth-like'' conditions on the planet and asserts that the most important factor to influence habitability is the presence of water in it's liquid state since ``all organisms with which we are familiar require liquid water during at least part of their life cycle" \citep{Kasting:1993p6301}. However, as our understanding of exoplanets and their ranges in sizes, masses, potential compositions, and stellar hosts has expanded, as a result of the Kepler Mission \citep{Batalha13}, so too must the definition of habitability \citep{Tasker17}. It is still of paramount importance that a planet maintain a stable temperature that is conducive to Earth-life. However, atmosphere is an important consideration -- for example, Venus-analogs which would not be considered habitable \citep{Kane14}. Additionally, the geochemical activity of a planet, namely plate tectonics or other recycling processes, climate, and geodynamo, must also be taken into consideration \citep{Foley16}. Ultimately, the term ``habitable" must be be expanded to not only consider if water could be in its liquid state on the surface of the planet, but whether the geochemical processes necessary  for life are present at all. It is only by applying a truly holistic approach, modeling both the chemical and physical properties along with interior and exterior cycling of a planetary system, that a planet can be considered ``alive" or even ``Earth-like.''

In this paper, we started by considering 10 stellar systems that are near to the Sun, namely stars around which it is most likely a terrestrial, rocky exoplanet will soon be discovered. When determining the mineralogies and structures of these potential planets, we realized that they were remarkably similar (see Paper II, \citealt{UnterbornHinkel}). The reason for the model homogeneity was due in part to the fact that, while the stellar abundances were notably disparate (see Table \ref{abunds1}-\ref{abunds2}), the molar fractions were consistent (Table \ref{molarerr} and Figure \ref{SiMg}). Additionally, when using the planet building code ExoPlex, we took advantage of the spread in the stellar abundance measurements -- or the range in determinations by different groups for the same element in the same star -- as the true uncertainty in the abundances. These two factors resulted in planetary mineralogies and structures that were relatively uniform \citep{UnterbornHinkel}.

Our primary assumption within this paper is that the composition of the host star is that of the resulting terrestrial planets. This assumption is well founded given the materials that go on to form rocky planets within the disk: high temperature, refractory condensates, primarily silicates and metallic iron. These refractory minerals are dominated by the so called major planet-building elements: O, Mg, Si, and Fe \citep[consequently the four most abundant elements in the Earth,][]{McD03}. While each of these elements are broadly considered refractory, they do not each condense at the exact same temperature. Instead, they condense at an initial temperature, continuing radially over a range of temperatures until reaching temperatures where they are entirely stable within solid rather than gas species \citep{bond_2010_aa,bond_2010_bb,Thiabaud15, Unterborn_2017}. The initial condensation temperatures of these refractory elements are within 3 K of each other, while the 50\% condensation temperatures are within 30 K of one another \citep{Lodders03}. Mixing between various compositionally distinct radial zones within the disk is possible, however, only small variations ($\sim$10\%) away from host start abundances and their associated ratios are expected \citep{bond_2010_aa, bond_2010_bb}. The minor planet-building elements, such as the moderate volatiles (Na, K), do not follow this one-to-one trend. This is due to their inherent volatility, where disk processes such as melting and subsequent impacts during formation fractionate their abundances relative to the major elements. While the moderate volatile elements are important for the potential crustal composition of a rocky exoplanet \citep{Unterborn_2017}, they do not drastically affect the bulk mantle mineralogy of the planet, and therefore variations in minor element abundance will not be reflected in mass-radius studies \citep{Dorn15, Unterborn_2017}. Thus, our assumption that host star composition is roughly that of the terrestrial planet is sound for our purposes of gauging the potential first-order mineralogy of these systems for broad comparative planetology.

While it is our intention to analyze the planetary structure for stars and planets very much unlike the Earth -- in terms of molar fractions, we leave that analysis to another paper. Instead, we focused on calculating what uncertainty (or spread) is needed in the stellar abundances in order to determine two populations of planets with mineralogies and structures that were unique and discernible. For our sample of 10 stars nearest to the Sun, we found that the precisions in the abundances need to be [Fe/H] $<$ 0.02 dex, [Si/H] $<$ 0.01 dex, [Al/H] $<$ 0.002 dex, while [Mg/H] and [Ca/H] $<$ 0.001 dex. Note that since all of the molar fractions were with respect to Mg, there is a degeneracy for the precision required for [Mg/H]. However, since Ca/Mg required the most strict precision, we adopted that requirement for [Mg/H] as well. For all of the stars within Hypatia, the precisions in the abundances were not much better than our smaller sub-sample: [Fe/H] and [Si/H] $<$ 0.03 dex, [Al/H] $<$ 0.004 dex, while [Mg/H] and [Ca/H] $<$ 0.002 dex. 

While the precision levels required to determine unique planetary structures are high, they aren't impossible. Ultimately, they require a more concerted effort on the part of both individual groups and the stellar abundance community as a whole to reduce both the error {\it and} the spread in the stellar abundances. Fortunately, within the last few years, there appears to be a more unified endeavor to better understand the measurement techniques between stellar abundance groups, for example the teams that participated in such studies as \citet{Smiljanic14, Jofre15, Hinkel16}.

However, even when stellar abundance compositions are measured consistently with high precision, the planetary interior and structure is only one aspect of habitability. The atmosphere, surface temperature, and activity levels of the host star need to be taken into account, such that the planetary surface and interior processes can be understood as a unified system. Therefore, we cannot say whether or not terrestrial planets orbiting the 10 stars nearest to the Sun, analyzed here, are habitable. However, given that many of the stellar hosts have low activity levels and that mineralogies are not too dissimilar to the Earth, we find that these 10 systems are a good place to look for \textit{potentially} habitable, rocky planets.

\appendix \label{appendix}
In an effort to be transparent with respect to stellar systems we did not analyze in this manuscript, below we list the particular systems we excluded and why, as noted in in the text.

\vspace{3mm}
\noindent
Stars removed because they are gravitational or spectroscopic binaries: HIP 71683 and 71681 ($\alpha$ Cen A \& B, respectively), HIP 104214 and 104217 (61 Cyg A \& B, respectively), HIP 84405 (36 Oph), HIP 84478 (V* V2215 Oph), HIP 19849 (omi02 Eri), HIP 3821 ($\eta$ Cas), HIP 37279 ($\alpha$ CMi), HIP 88601 (70 Oph), HIP 99461 (HD 191498), HIP 73184 (HD 131977), HIP 12114 (HD 16160), HIP 5336 ($\mu$ Cas), HIP 113283 (Fomalhaut), HIP 88601 (70 Oph), HIP 73184 (HD 131977), HIP 99461 (HD 191408), HIP 12114 (HD 16160), HIP 86974 ($\mu$ Her), HIP 61317 ($\beta$ CVn), HIP 27913 ($\chi$01 Ori), HIP 32984 (HD 50281), HIP 84720 (41 Ara), HIP 99825 (HD 192310), HIP 27072 ($\gamma$ Lep).

\vspace{3mm}

\noindent
Stars removed because they are already known to host exoplanets: HIP 113020 (BD-15 6290), HIP 16537 ($\epsilon$ Eri), HIP 15510 (82 Eri), HIP 8102 ($\tau$ Ceti), and HIP 64924 (61 Vir) from our sample list.

\vspace{3mm}

\noindent
Stars excluded because they had abundances values for any of the 5 elements [X/H] with spreads $>$ 0.70 dex: HIP 45343 (HD 79210), HIP 49908 (HD 88230), HIP 85295 (GJ 673), HIP 113576 (HD 217357), and HIP 1599 ($\zeta$ Tuc).

\clearpage
\LongTables
\begin{deluxetable*}{ccccccccc}
  \tablecaption{\label{phys} Measured Stellar Properties of the Sample}
  \tablehead{
\colhead{HIP } &  \colhead{Alt.} & \colhead{ RA } &  \colhead{ Dec } &  \colhead{ dist } &  \colhead{ Spec } &  \colhead{ V } &  \colhead{ \teff } &  \colhead{ \lg  } \\
\colhead{ } & \colhead{Name } &  \colhead{ (deg) } &  \colhead{ (deg) } &  \colhead{ (pc) } &  \colhead{ Type } &  \colhead{ mag } &  \colhead{(K) } &  \colhead{ } }
108870 & $\epsilon$ Indi& 330.840 & -56.786 & 3.62 & K4V(k) & 4.69 & 4638.5 & 4.54  \\ 
96100 & $\sigma$ Dra & 293.090 & \,\,69.661 & 5.76 & K0V & 4.68 & 5278.7 & 4.49 \\ 
99240 & $\delta$ Pav&  302.182 & -66.182 & 6.11 & G8IV & 3.56 & 5546.1 & 4.26 \\ 
3765 & HD4628 & \,\,\,\,12.096 & \,\,\,\,\,5.281 & 7.45 & K2.5V & 5.74 & 4998.2 & 4.63  \\ 
2021 & $\beta$ Hyi & \,\,\,\,\,6.438 & -77.254 & 7.46 & G0V & 2.79 & 5799.0 & 4.02   \\ 
7981 & 107 Psc & \,\,\,25.624 & \,\,20.269 & 7.53 & K1V & 5.24 & 5197.3 & 4.45  \\
23311 & HD32147 & \,\,\,75.204 & \,\,\,\,-5.754 & 8.71 & K3+V & 6.21 & 4808.7 & 4.41   \\
17378 & $\delta$ Eri & \,\,\,55.812 & \,\,\,\,-9.763 & 9.04 & K1III-IV & 3.54 & 5029.1 & 3.82 \\
57939 & HD103095 & 178.245 & \,\,\,\,37.719 & 9.09 & K1VFe-1.5 & 6.45 & 5041.2 & 4.63   \\
64394 & $\beta$ Com & 197.968 & \,\,\,\,27.878 & 9.13 & F9.5V & 4.25 & 5975.5 & 4.43  
 \enddata
 \end{deluxetable*}

\begin{deluxetable*}{p{0.7cm}p{0.4cm}p{0.4cm}p{0.4cm}p{0.5cm}p{0.5cm}p{0.4cm}p{0.4cm}p{0.5cm}p{0.4cm}p{0.5cm}p{0.5cm}p{0.4cm}p{0.4cm}} 
  \tablecaption{\label{physHZ}{Calculated Stellar Properties of the Sample}}
  \tablehead{
\colhead{HIP } &    \colhead{Calc St.}  & \multicolumn{3}{c}{Optimistic Inner} & \multicolumn{3}{c}{Conservative Inner} & \multicolumn{3}{c}{Conservative Outer} & \multicolumn{3}{c}{Optimistic Outer}
\\
\colhead{ } &  \colhead{Mass} & \colhead{HZ } & \colhead{ RV amp  } &  \colhead{ Period  } & \colhead{ HZ } & \colhead{ RV amp } &  \colhead{ Period } & \colhead{ HZ } & \colhead{ RV amp } &  \colhead{ Period  } & \colhead{ HZ } & \colhead{ RV amp } &  \colhead{ Period } 
\\
\colhead{ } &  \colhead{(M$_{\odot}$)} & \colhead{(AU) } & \colhead{ (m/s) } &  \colhead{  (days) } & \colhead{ (AU)} & \colhead{  (m/s) } &  \colhead{ (days) } & \colhead{  (AU)} & \colhead{(m/s) } &  \colhead{  (days) } & \colhead{  (AU)} & \colhead{ (m/s) } &  \colhead{  (days) } 
}
108870 &   0.704 & 0.796 & 0.1699 & 309.18 & 1.008 & 0.1509 & 440.59 & 1.851 & 0.1114 & 1096.37 & 1.950 & 0.1085 & 1185.49 \\
96100 &   0.865 & 0.772 & 0.1265 & 266.29 & 0.978 & 0.1124 & 379.70 & 1.751 & 0.0840 & 909.61 & 1.840 & 0.0819 & 979.84 \\
99240 &  0.937 & 0.761 & 0.1131 & 250.52 & 0.964 & 0.1005 & 357.17 & 1.710 & 0.0755 & 843.83 & 1.802 & 0.0735 & 912.83 \\
3765 &   0.793 & 0.783 & 0.1432 & 284.15 & 0.992 & 0.1272 & 405.20 & 1.796 & 0.0945 & 987.09 & 1.893 & 0.0921 & 1068.13 \\
2021 & 1.006 & 0.750 & 0.1024 & 236.51 & 0.949 & 0.0910 & 336.64 & 1.674 & 0.0685 & 788.68 & 1.762 & 0.0668 & 851.68 \\
7981 &  0.844 & 0.776 & 0.1309 & 271.72 & 0.982 & 0.1164 & 386.81 & 1.765 & 0.0868 & 932.06 & 1.860 & 0.0846 & 1008.31 \\
23311 &  0.745 & 0.790 & 0.1564 & 297.01 & 1.001 & 0.1389 & 423.62 & 1.826 & 0.1029 & 1043.70 & 1.925 & 0.1002 & 1129.72 \\
17378 &   0.801 & 0.782 & 0.1412 & 282.21 & 0.991 & 0.1254 & 402.59 & 1.790 & 0.0933 & 977.32 & 1.886 & 0.0909 & 1056.98 \\
57939 &   0.804 & 0.782 & 0.1403 & 281.66 & 0.990 & 0.1247 & 401.21 & 1.790 & 0.0928 & 975.44 & 1.883 & 0.0904 & 1052.44 \\
64394 &   1.055 & 0.741 & 0.0959 & 226.77 & 0.939 & 0.0852 & 323.48 & 1.648 & 0.0643 & 752.12 & 1.736 & 0.0626 & 813.16 \\  
\enddata
\end{deluxetable*}

\LongTables
\begin{deluxetable*}{ccccccccccc}
  \tablecaption{\label{abunds1} Stellar Abundances for the 10 Nearest Stars}
  \tablehead{
\colhead{HIP } &  \colhead{ [Mg/H] } &  \colhead{ sp[Mg/H] } &  \colhead{ [Al/H] } &  \colhead{ sp[Al/H] } &  \colhead{ [Mg/H]$_\text{Al}$ } &  \colhead{ sp[Mg/H]$_\text{Al}$ } &  \colhead{ [Si/H] } &  \colhead{ sp[Si/H]} &  \colhead{ [Mg/H]$_\text{Si}$ } &  \colhead{ sp[Mg/H]$_\text{Si}$ } 
  }
 \startdata
108870 & -0.08 & 0.37 & -0.04 & 0.01 & -0.08 & 0.37 & -0.12 & 0.20 & -0.08 & 0.37 \\ 
96100 & -0.11 & 0.31 & -0.17 & 0.27 & -0.03 & 0.31 & -0.20 & 0.21 & -0.11 & 0.31 \\ 
99240 & 0.40 & 0.16 & 0.38 & 0.27 & 0.41 & 0.16 & 0.39 & 0.15 & 0.40 & 0.16 \\ 
3765 & -0.09 & 0.26 & -0.15 & 0.24 & -0.03 & 0.26 & -0.20 & 0.10 & -0.09 & 0.26 \\ 
2021 & -0.01 & 0.18 & -0.06 & 0.27 & 0.01 & 0.14 & -0.06 & 0.16 & -0.01 & 0.18 \\ 
7981 & 0.10 & 0.15 & 0.03 & 0.19 & 0.11 & 0.15 & 0.01 & 0.15 & 0.10 & 0.15 \\ 
23311 & 0.55 & 0.43 & 0.49 & 0.36 & 0.45 & 0.43 & 0.40 & 0.28 & 0.55 & 0.43 \\ 
17378 & 0.28 & 0.32 & 0.26 & 0.24 & 0.28 & 0.32 & 0.18 & 0.25 & 0.28 & 0.32 \\ 
57939 & -1.02 & 0.27 & -1.42 & 0.25 & -1.13 & 0.27 & -1.09 & 0.22 & -1.02 & 0.27 \\ 
64394 & 0.06 & 0.23 & 0.05 & 0.13 & 0.06 & 0.08 & 0.03 & 0.13 & 0.06 & 0.23 
 \enddata
 \end{deluxetable*}

\LongTables
\begin{deluxetable*}{ccccccccc}
  \tablecaption{\label{abunds2} Stellar Abundances for the 10 Nearest Stars (cont.)}
  \tablehead{
\colhead{HIP } &  \colhead{ [Ca/H] } &  \colhead{ sp[Ca/H] } &  \colhead{ [Mg/H]$_\text{Ca}$ } &  \colhead{ sp[Mg/H]$_\text{Ca}$ } &  \colhead{ [Fe/H] } &  \colhead{ sp[Fe/H] } &  \colhead{ [Mg/H]$_\text{Fe}$ } &  \colhead{ sp[Mg/H]$_\text{Fe}$}
  }
 \startdata
108870 & -0.08 & 0.48 & -0.08 & 0.37 & -0.07 & 0.41 & -0.08 & 0.37 \\ 
96100 & -0.16 & 0.22 & -0.07 & 0.31 & -0.15 & 0.35 & -0.11 & 0.31 \\ 
99240 & 0.32 & 0.24 & 0.41 & 0.10 & 0.37 & 0.27 & 0.40 & 0.16 \\ 
3765 & -0.18 & 0.21 & -0.03 & 0.26 & -0.20 & 0.29 & -0.09 & 0.26 \\ 
2021 & 0.00 & 0.02 & 0.03 & 0.08 & -0.09 & 0.49 & -0.01 & 0.18 \\ 
7981 & -0.01 & 0.17 & 0.11 & 0.15 & 0.01 & 0.35 & 0.10 & 0.15 \\ 
23311 & 0.32 & 0.36 & 0.45 & 0.43 & 0.32 & 0.26 & 0.55 & 0.43 \\ 
17378 & 0.14 & 0.30 & 0.28 & 0.32 & 0.19 & 0.44 & 0.28 & 0.32 \\ 
57939 & -1.05 & 0.17 & -1.01 & 0.27 & -1.24 & 0.27 & -1.02 & 0.27 \\ 
64394 & 0.11 & 0.16 & 0.07 & 0.08 & 0.11 & 0.33 & 0.06 & 0.23  
 \enddata
 \end{deluxetable*}

\LongTables
\begin{deluxetable*}{ccccccccc}
  \tablecaption{\label{ratios} Molar Ratios and Errors of Sample Stars (dex)}
  \tablehead{
\colhead{HIP } &  \colhead{ Al/Mg } &  \colhead{ $\sigma_{Al/Mg}$} &  \colhead{ Si/Mg } & \colhead{ $\sigma_{Si/Mg}$ } &  \colhead{ Ca/Mg } &  \colhead{ $\sigma_{Ca/Mg}$ } &  \colhead{ Fe/Mg } &  \colhead{ $\sigma_{Fe/Mg}$ } 
  }
 \startdata 
108870 & 0.09 & 0.43 & 0.87 & 0.48 & 0.06 & 0.70 & 0.83 & 0.64 \\
96100 & 0.06 & 0.47 & 0.78 & 0.43 & 0.05 & 0.44 & 0.74 & 0.54 \\
99240 & 0.08 & 0.36 & 0.93 & 0.25 & 0.05 & 0.30 & 0.76 & 0.36 \\
3765 & 0.06 & 0.41 & 0.74 & 0.32 & 0.04 & 0.38 & 0.63 & 0.45 \\
2021 & 0.07 & 0.35 & 0.85 & 0.28 & 0.06 & 0.09 & 0.68 & 0.60 \\
7981 & 0.07 & 0.28 & 0.78 & 0.24 & 0.05 & 0.26 & 0.66 & 0.44 \\
23311 & 0.09 & 0.65 & 0.68 & 0.59 & 0.05 & 0.65 & 0.48 & 0.58 \\
17378 & 0.08 & 0.46 & 0.76 & 0.47 & 0.04 & 0.51 & 0.66 & 0.63 \\
57939 & 0.04 & 0.42 & 0.81 & 0.40 & 0.06 & 0.37 & 0.49 & 0.44 \\
64394 & 0.08 & 0.18 & 0.89 & 0.30 & 0.07 & 0.21 & 0.91 & 0.46 \\
\\
Sun & 0.09 &  & 0.95 &  & 0.06 &  & 0.81 &  
 \enddata
 \end{deluxetable*}

\section*{Acknowledgments}
The authors would like to thank Eric Mamajek and Maggie Turnbull for fruitful conversations as well as Stephen Kane for helpful calculations. NRH would like to thank CHW3 and acknowledges the support of the Vanderbilt Office of the Provost through the Vanderbilt Initiative in Data-intensive Astrophysics (VIDA) fellowship.
The research shown here acknowledges use of the Hypatia Catalog Database, an online compilation of stellar abundance data as described in \citet{Hinkel14}, which was supported by NASA's Nexus for Solar System Science (NExSS) research coordination network and VIDA.
The results reported herein benefited from collaborations and/or
information exchange within NASA's NExSS research coordination network sponsored by NASA's Science
Mission Directorate.
This research has made use of the NASA Exoplanet Archive, which is operated by the California Institute of Technology, under contract with the National Aeronautics and Space Administration under the Exoplanet Exploration Program. 
This research has made use of the Vizier catalogue access tool and Simbad portal via the CDS, Strasbourg, France. 

 \newcommand{\noop}[1]{}

\end{document}